# New Southern Galaxies With Active Nuclei. Part II. *

M. A. G. Maia$^{a}$, J. A. Suzuki$^{a}$, L. N. da Costa$^{a,b,**}$, C. N. A. Willmer$^{a,c}$ and C. Rité$^{a}$

$^{a}$ Departamento de Astronomia, Observatório Nacional, Rua Gal. José Cristino 77, Rio de Janeiro - 20921 - RJ, Brazil.
$^{b}$ European Southern Observatory, Karl-Schwarzschild-Strasse 2, 85748 Garching bei München, Germany.
$^{c}$ Institut d'Astrophysique, 98 bis, Boulevard Arago, F-75014 Paris, France.
E-mail: maia@on.br, suzuki@on.br, ldacosta@eso.org, cnaw@on.br, rite@on.br

**Abstract.** This paper contains a list of new AGN candidates identified from the examination of 3500 optical spectra contained in the database of the Southern Sky Redshift Survey Extension (SSRS2). The classification of galaxies was done using standard diagnostics and a total of (5) Seyfert 1, (12) Seyfert 2 and (10) LINERs were found. We also present a list of 60 galaxies for which we could not secure a definite classification, but which might present some level of nuclear activity.

## 1. Introduction

A considerable effort has been undertaken to understand the physical processes generating the phenomena observed in active galactic nuclei (AGN), where a vast amount of energy is produced on short timescales in a small volume of the parent galaxy. In order to understand the underlying physical mechanism responsible for the phenomenon, different observational techniques must be used such as time monitoring and multiwavelength studies. Equally important to explain these phenomena, is the availability of large number of observed objects covering a wide range in redshift, local density regimes and levels of nuclear activity, given the great variety of behavior presented by active galaxies. This fact provides an important motivation for a systematic search for new candidates. In addition to the surveys especially designed to find new AGN candidates, active galaxies can also be identified as a by-product of the various ongoing redshift surveys of galaxies, where several thousands of spectra are accumulated in a systematic fashion following well defined selection criteria. An additional advantage of this procedure is that the relation of the AGNs with their surrounding environment may be established. Because these redshift surveys are directed at studying the properties of galaxies and structures of galaxies on large scales, they should allow us to determine some of the statistical properties of AGNs, as the parent samples are relatively unbiased towards this kind of object. One of these properties is, for instance, the population density of active galaxies relative to normal ones, which is still not well known, but which might hold some important clues on the evolution of galaxies, their contribution to the X-ray background and to evaluate the possibility of using AGNs as tracers of galaxy distributions (e.g. Huchra and Burg 1992; Osterbrock and Martel 1993).

In this paper we report the results of a search for galaxies which could be hosts of active nuclei, using the database of the Southern Sky Redshift Survey, SSRS (e.g., da Costa et al. 1988, 1991) as well as its extensions (Fairall et al. 1992; Huchra et al. 1993), and SSRS2 (da Costa et al. 1994). The fact that this survey is not specifically aimed at measuring AGNs has the appealing point that it allows an easier detection of low-luminosity AGNs relative to the conventional low-dispersion objective-prism surveys. This paper is a follow up of the work by Maia et al. (1987), where the first SSRS AGN candidates were catalogued, and provides the groundwork for a study of the spatial distribution of active nuclei in the SSRS survey. The data reported here have been collected during the last 6 years and consist mainly of galaxies in both galactic caps in the region $|b| > 30$ and $\delta < 0$. We discuss the observational procedures in section 2, and present the results in section 3.

## 2. Observations

The observations reported here were made at four different sites. The data collected at the 1.5 m telescope of the Laboratório Nacional de Astrofísica (LNA), Brasópolis, Brazil, used the Observatório Nacional intensified photon-counting Reticon detector described by da Costa et al. (1984), who also discuss the reduction procedures. A 6Å mean resolution was obtained for a 4700-7100Å spectral range. This detector was also used on the 2.15 m telescope of the Complejo Astronomico El Leoncito (CASLEO), San Juan, Argentina, but having a slightly larger spectral coverage (4450-7100Å). The remaining data were collected at the Cerro Tololo Interamerican Observatory (CTIO) 1.5m telescope using a CCD with 421 x 576 pixels and the European Southern Observatory (ESO) Spectroscopic 1.52m telescope, with a Ford CCD with 2048 x 2048 pixels. The spectral coverage depends on the epoch, but in general was in the range 3900-7100Å for ESO and 4800-7000Å for CTIO. The CTIO mean resolution was about 8Å, while for ESO data it is $\approx 4$Å.

As the survey was designed to measure redshifts of galaxies, most of the emission-line spectra have a low signal-to-noise ratio in the continuum, particularly in the case of data obtained with the intensified reticon, as it allowed examination of the spectra in real-time during the acquisition process. Nevertheless, for the majority of galaxies the quality of the spectra is suitable for the identification of features characteristic of AGNs. We should also note that because of time constraints spectrophotometric standards where not observed systematically, and, as a result, very few of our spectra can be flux calibrated. This is particularly important for data obtained with the intensified reticon. However, as the aim of this work is to provide a *list* of AGN *candidates* for future and more detailed work, these limitations should not be excessively restrictive.

## 3. Results

About 3500 new spectra in the SSRS database were visually examined and an attempt was made to separate those with emission-lines features typical of H II regions from those with AGN characteristics. We have used for this purpose the diagnostics proposed by Baldwin, Phillips and Terlevich (1981) who use the ratio of forbidden to permitted line intensities to delineate the regions occupied by conventional H II regions; by low-ionization nuclear-emission regions (LINERs, Heckman 1980); or Seyfert-like activity. Although different combinations of line intensities may be used, for our spectra uncalibrated in flux it is convenient to use lines which are close to each other. Therefore AGN candidates were selected primarily on the basis of the ratio between the equivalent widths of the contiguous [N II]$\lambda 6583$ and H$\alpha$.

Spectra with ratio $R_1 \equiv$ [ N II ]$\lambda 6583$ / H$\alpha > 0.7$ were assumed to be either Seyfert or LINERs. The separation between these two classes was made whenever possible based on the measured ratio $R_2 \equiv$ [ O III ]$\lambda 5007$ / H$\beta$ and following the criteria that galaxies with $R_2 < 3$ were LINERs and those with $R_2 > 3$ were Seyferts. The lack of blue sensitivity makes this diagnostic less reliable and sometimes subjective.

Our classification should be regarded as tentative also because no corrections were applied to the spectra due to line blending, interstellar reddening and the underlying stellar absorption. For instance, in some galaxies the presence of strong Balmer absorption will conceal the H$\beta$ emission-line in the spectra, making it difficult to separate a LINER from a Seyfert 2 spectrum. One should also have in mind that the classical definition of LINERs does not immediately imply an AGN-like activity as it has been shown that LINERs with weak [ O I ]$\lambda 6300$ relative to H$\alpha$ ($< 1/6$) can be accounted for by a photoionization model (Filippenko and Terlevich 1992).

Taking into consideration the above remarks we have adopted the following classification scheme: 1) S1 - objects with broad Balmer lines typical of Seyfert 1 galaxies; 2) S2 - galaxies with $R_1 > 0.7$ and $R_2 > 3$; 3) Li - galaxies with $R_1 > 0.7$ and $R_2 < 3$; 4) SL: - galaxies with visible [ O III ]$\lambda 5007$ lines but undetected H$\beta$; 5) SL? - galaxies with $R_1 > 0.7$ but no visible [ O III ]$\lambda 5007$ and H$\beta$; 6) L? - galaxies with $R_1 \sim 0.7$ and $R_2 < 3$. To further aid the reader, for each galaxy we point out in the tables the possible presence of H$\beta$ in absorption, the strength of [ O I ]$\lambda 6300$ line relative to H$\alpha$ and spectra with poor sky subtraction.

About 2% of the objects examined in the SSRS database presented characteristics of AGN galaxies. After a first selection, the literature was searched for any previous reference of nuclear activity of the candidate galaxies. For this purpose we used the 6$^{th}$ edition of *A Catalogue of Quasars and Active Nuclei* by Véron-Cetty and Véron (1993), as well as the literature covering the remaining period of the aforementioned catalogue up to December 1994. Finally, the NASA/IPAC Extragalactic Database was searched.

We have also carried out a search of the *IRAS Faint Source Catalog* to find possible infrared (IR) counterparts of the objects in our final list. For those with IR detection, the far-infrared luminosity was calculated. The IRAS counterparts were searched using an adapted version of a program for matches on *The Green Bank Sky Maps and Radio Source Catalog* CD-ROM, produced by NRAO. All the matches with values of M (maximum normalized position difference) smaller than 3.5 were accepted. The optical size of the galaxies and error ellipses of IRAS sources were taken into account. A significant fraction of the Seyfert galaxies have IR counterparts ($\approx 60\%$). They are not the typical ultraluminous IR galaxies ($L_{FIR} \geq 10^{12} L_\odot$), being instead low-luminosity IR objects ($L_{FIR} < 10^{11} L_\odot$).

The resulting list of candidates with no previous reference of AGN activity is presented in Table 1 (for CASLEO and LNA observations) and in Table 2 (for ESO and CTIO). The tables contain the following information:

Column (1) – The identification based on 1950.0 equatorial coordinates. For ESO galaxies we used the coordinates given by Lauberts and Valentijn (1989), for the remaining objects the coordinates of the STScI *Guide Star Catalog* (Lasker et al. 1990) precessed to Epoch 1950.0.

Column (2) – Galaxy identification in the ESO, NGC or MCG catalogues if available.

Columns (3) and (4) – Equatorial coordinates for epoch 1950.0 taken from Lauberts and Valentijn (1989), for objects in the ESO catalogue. For the non-stellar objects listed in the STScI *Guide Star Catalog* that were observed as part of the SSRS2, and which are not contained in the Lauberts and Valentijn (1989) catalogue, the published coordinates were precessed from Epoch 2000.0 to 1950.0. The precision in the coordinates is $\approx 3''$ for ESO galaxies (Lauberts 1989) and $\approx 1''$ for STScI (Lasker et al. 1990).

Column (5) and (6) – The apparent B magnitude, $m_B$, and its reference given by Lauberts and Valentijn (1989) for ESO galaxies (coded as 1), or derived from the STScI instrumental magnitudes to correspond to a blue magnitude measured within B=26 mag/arcsec$^2$ isophote according to Alonso et al. (1993), coded as "2" in column 6.

Column (7) – Morphological type quoted by Lauberts and Valentijn (1989) whenever available. For objects indicated with an asterisk, the morphological classification is based on a visual inspection of the field in the CD-ROMs of the STScI *Digitized Sky Survey*.

Column (8) – Observatory where the spectrum was obtained.

Column (9) – Heliocentric velocity, $V_\odot$, in km/s.

Column (10) – Absolute blue magnitude, $M_B$, derived from the Hubble distance of the galaxy using the measured redshift and $H_o = 75$ km s$^{-1}$ Mpc$^{-1}$.

Column (11) – Total far-infrared luminosity between $\sim 40 \mu$m and $\sim 120 \mu$m, $L_{FIR}$, in $L_\odot$, computed following Lonsdale et al. (1985):

$$L_{FIR} = 3.95 \times 10^5 (2.58 f_{60} + f_{100}) \times d^2,$$

where $f_{60}$ and $f_{100}$ are the 60$\mu$m and 100$\mu$m infrared flux densities, in Jy, from the *IRAS Faint Source Catalog* and $d$, is the distance in Mpc.

Columns (12) and (13) – $R_1$ and $R_2$ as defined in the text.

Column (14) – Emission type, where the different symbols used to identify the type of emission are as discussed above.

Column (15) – Comments to the table where (1) indicates evidence of strong $H\beta$ absorption; (2) and (3) indicate the estimated strength of the [OI] $\lambda 6300$ line relative to $H\alpha$ with (2) denoting weak [OI] $\lambda 6300$ line (roughly $< 1/6$ of the $H\alpha$ line) and (3) strong [OI] $\lambda 6300$; (4) poor sky subtraction.

The spectra for galaxies listed in tables 1 and 2 are displayed in figs. 1 and 2 respectively, with fluxes in arbitrary units. For a few spectra it was possible to use a spectrophotometric standard star to remove the detector response. These objects are identified in the figures by an asterisk next to their names. For the remaining galaxies, we tried to remove the instrumental response by fitting the continuum and dividing the spectrum by the fit. The spectra show a large range of signal to noise ratios (S/N).

The modest FIR emission of objects in our catalogue seems to be compatible (at least in the case of Seyferts), with the fact that the nuclear activity is also modest. This may be a consequence of the selection criteria used to generate the sample of objects (diameter or magnitude limited sample), instead of characteristics that are more common to AGNs like bright nucleus, prominent emission in the blue band, among others. So far we have examined about 5000 spectra including the previous work of Maia et al. (1987). The percentual proportion of AGN galaxies in the SSRS sample is approximately 2%, and will be the subject of a study of the AGN phenomena in that sample which we expect to report in the future.

We acknowledge the scientific and technical personnel involved in the SSRS effort for their valuable help, as well as the staff of LNA, CASLEO, CTIO and ESO for their assistance during the observations. We thank the referee, Dr. M.P. Véron-Cetty, whose suggestions have helped to improve this paper. This work was supported in part by CNPq grants 800745/91-4 and 301366/86-1 (MAGM), 104595/92-2 and PIBIC/ON (JAS). This research has made use of the NASA/IPAC Extragalactic Database (NED) which is operated by the Jet Propulsion Laboratory, Caltech, under contract with the National Aeronautics and Space Administration.

## References

Alonso, M.V., da Costa, L.A.N., Pellegrini, P.S., and Kurtz, M.J. 1993, AJ 106, 676

Baldwin, J., Phillips, M., and Terlevich, R. 1981, PASP 93, 5

da Costa, L.N., Geller, M.J., Pellegrini, P.S., Latham, D.W., Fairall, A.P., Mazke, R.O., Willmer, C.N.A., Huchra, J.P., Calderon, J.H., Ramella, M., and Kurtz, M.J. 1994, ApJ 424, L1

da Costa, L.N., Pellegrini, P.S., Davis, M, Meiksin, A., Sargent, W.L.W., and Tonry, J. 1991, ApJS 76, 935

da Costa, L.N., Pellegrini, P.S., Nunes, M.A., Willmer, C., and Latham, D.W. 1984, AJ 89, 1310

da Costa, L.N., Pellegrini, P.S., Sargent, W.L.W, Tonry, J., Davis, M., Meiksin, A., Latham, D.W., Menzies, J.W., and Coulson, I.A. 1988, ApJ 327, 544

Fairall, A., Willmer, C., Calderón, J.H., Latham, D.W., da Costa, L.N., Pellegrini, P.S., Nunes, M.A., Focardi, P., and Vettolani, G. 1992, AJ, 103, 11

Filippenko, A.V., and Terlevich, R. 1992, ApJ 397, L79

Heckman, T.M. 1980, A&A 87, 152

Huchra, J., Latham, D.W., da Costa, L.N., Pellegrini, P.S., and Willmer, C.N.A. 1993, AJ 105, 1637

Huchra, J.P., and Burg, R. 1992, ApJ 393, 90

Lasker, B.M., Sturch, C.R., McLean, B.M., Russel, J.L., Jenker, H., and Shara, M. 1990, AJ 99, 2019

Lauberts, A. 1982, The ESO/Uppsala Survey of the ESO(B) Atlas, (European Southern Observatory, Garching-bei-München.)

Lauberts, A., and Valentijn, E.A. 1989, The Surface Photometry Catalogue of The ESO-Uppsala Galaxies, (European Southern Observatory, Garching-bei-München.)

Lonsdale, C.J., Helou, G., Good, J.C., and Rice, W.L. 1985, Cataloged Galaxies and Quasars Observed in the IRAS Survey (Jet Propulsion Laboratory, Pasadena)

Maia, M.A.G., da Costa, L.N., Willmer, C., Pellegrini, P.S., and Rité, C. 1987, AJ 93, 546

Osterbrock, D.E., and Martel, A. 1993, ApJ 414, 552.

Véron-Cetty, M.-P., and Véron P. 1993, A Catalogue of Quasars and Active Nuclei, $6^{th}$ Edition. (Garching bei München: European Southern Observatory)

### Figure Captions

**Figure 1.** Spectra of AGN candidates observed at CASLEO and LNA. Fluxes are in arbitrary units.

**Figure 2.** Spectra of AGN candidates observed at CTIO and ESO.

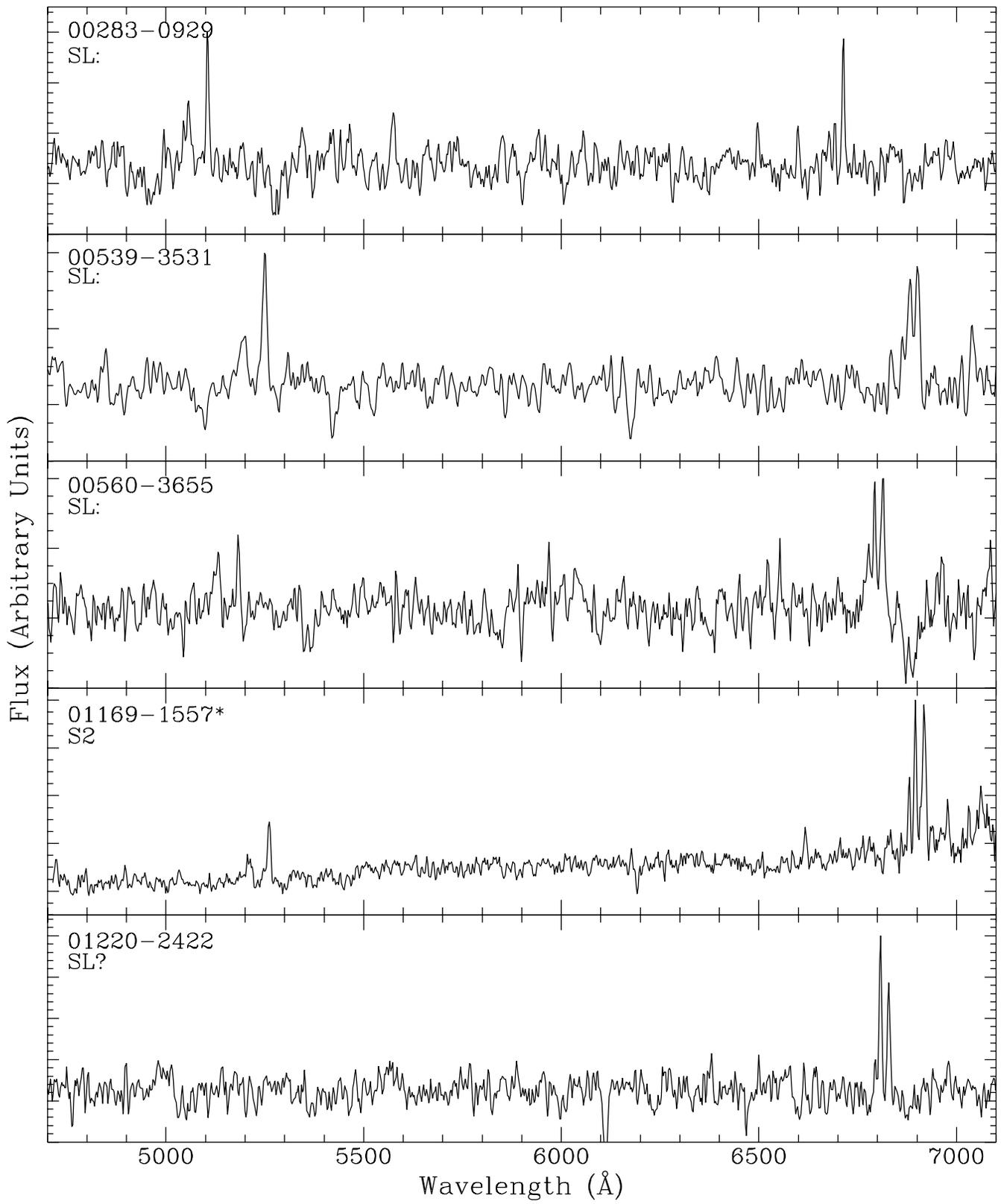

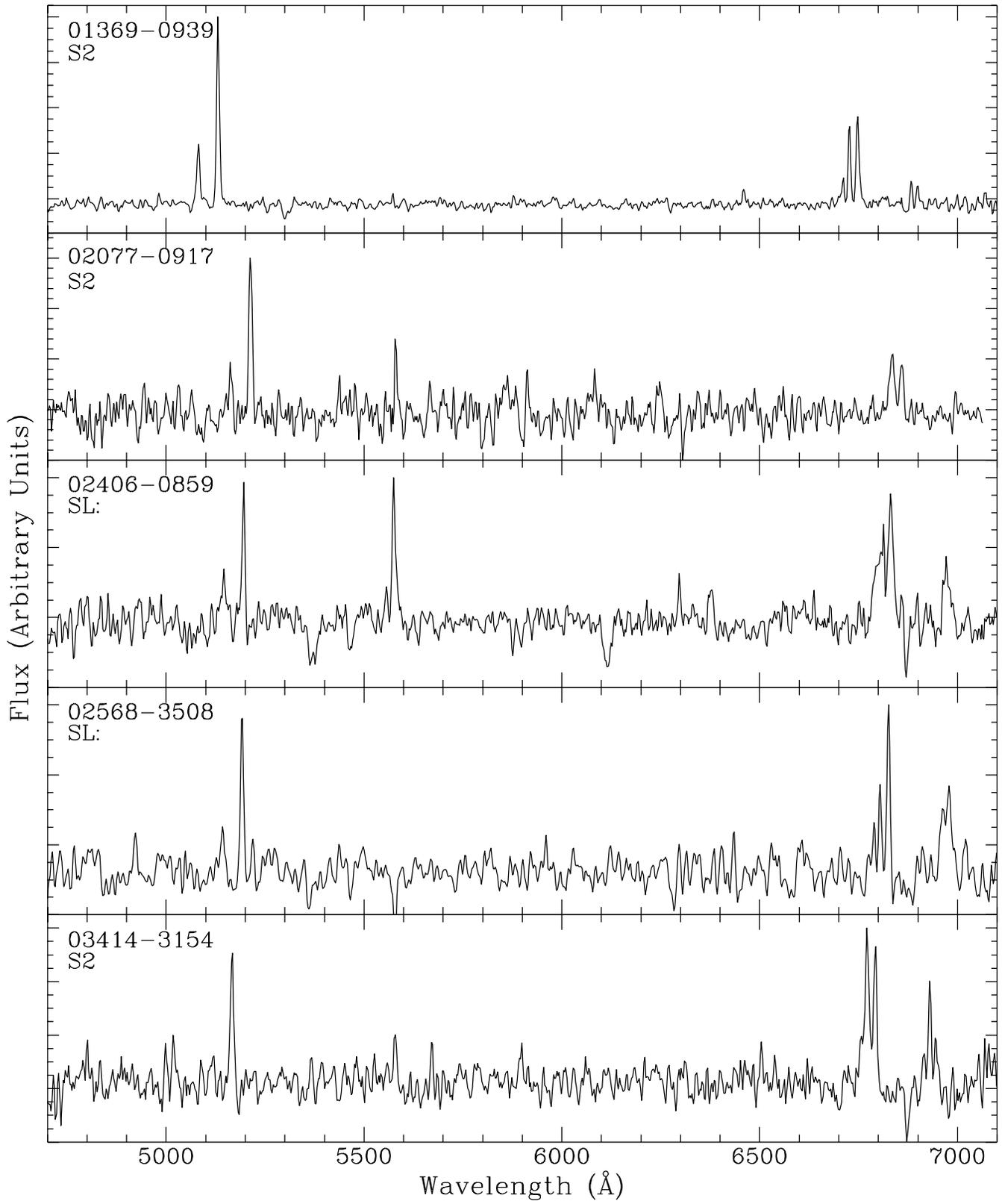

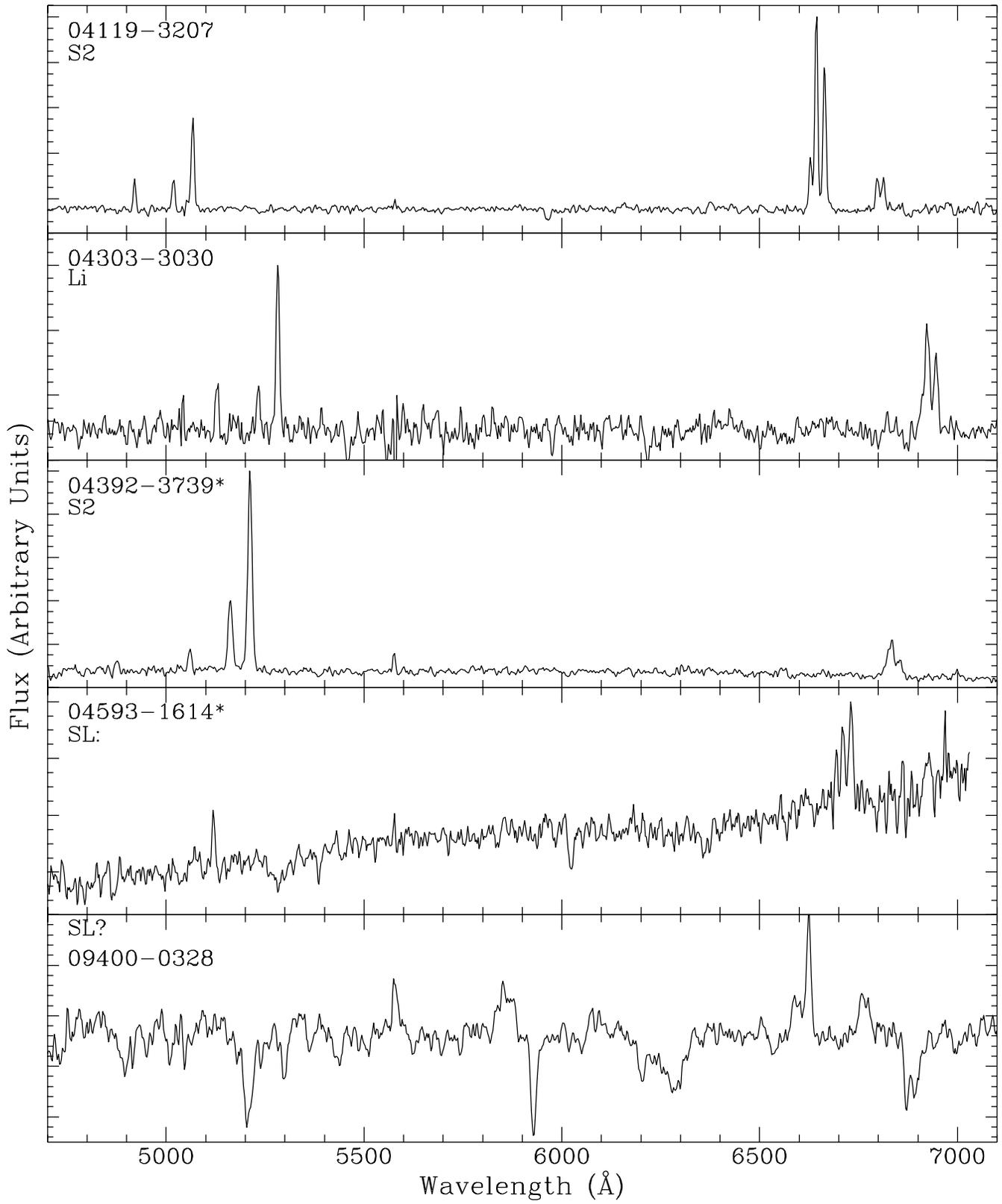

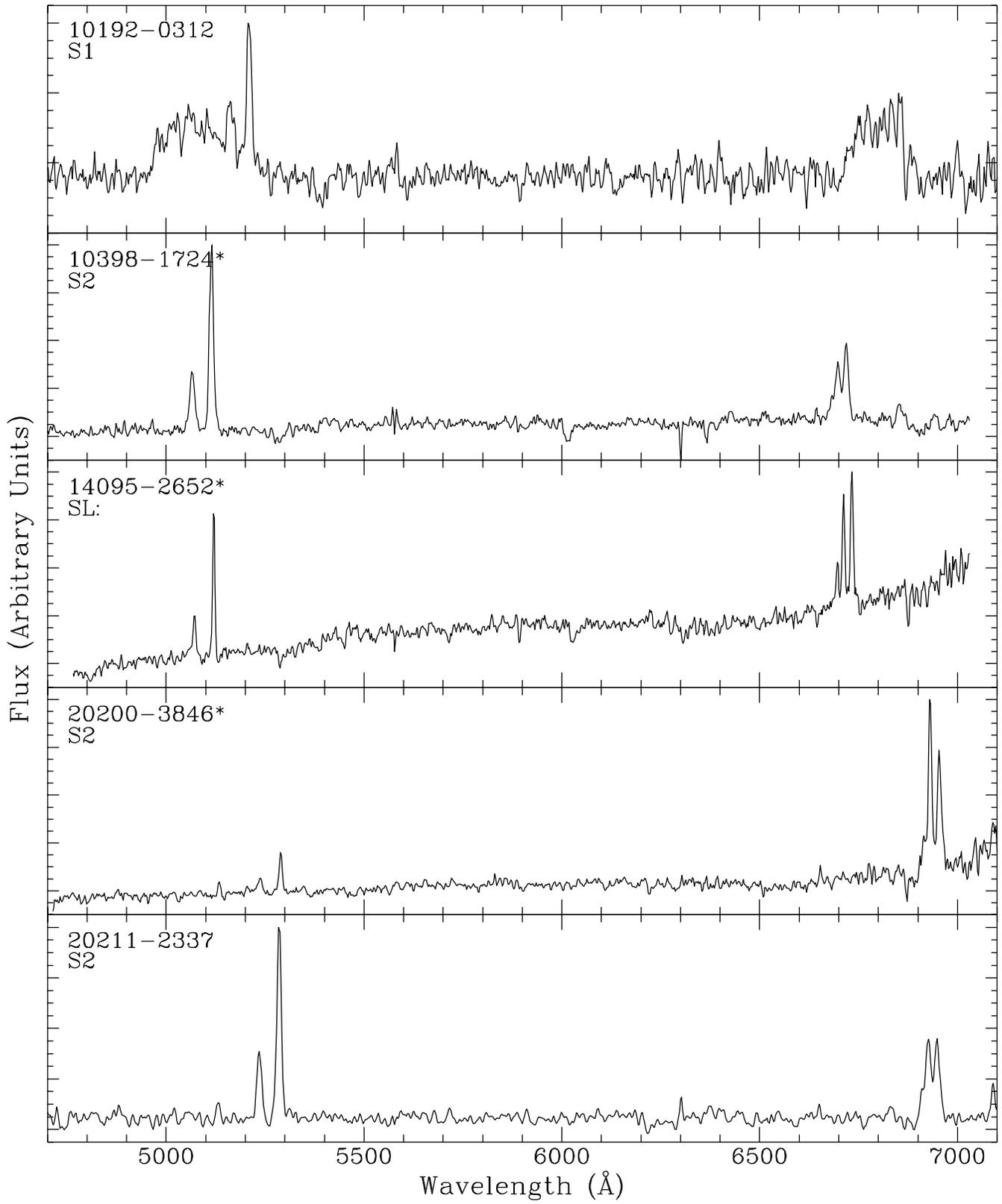

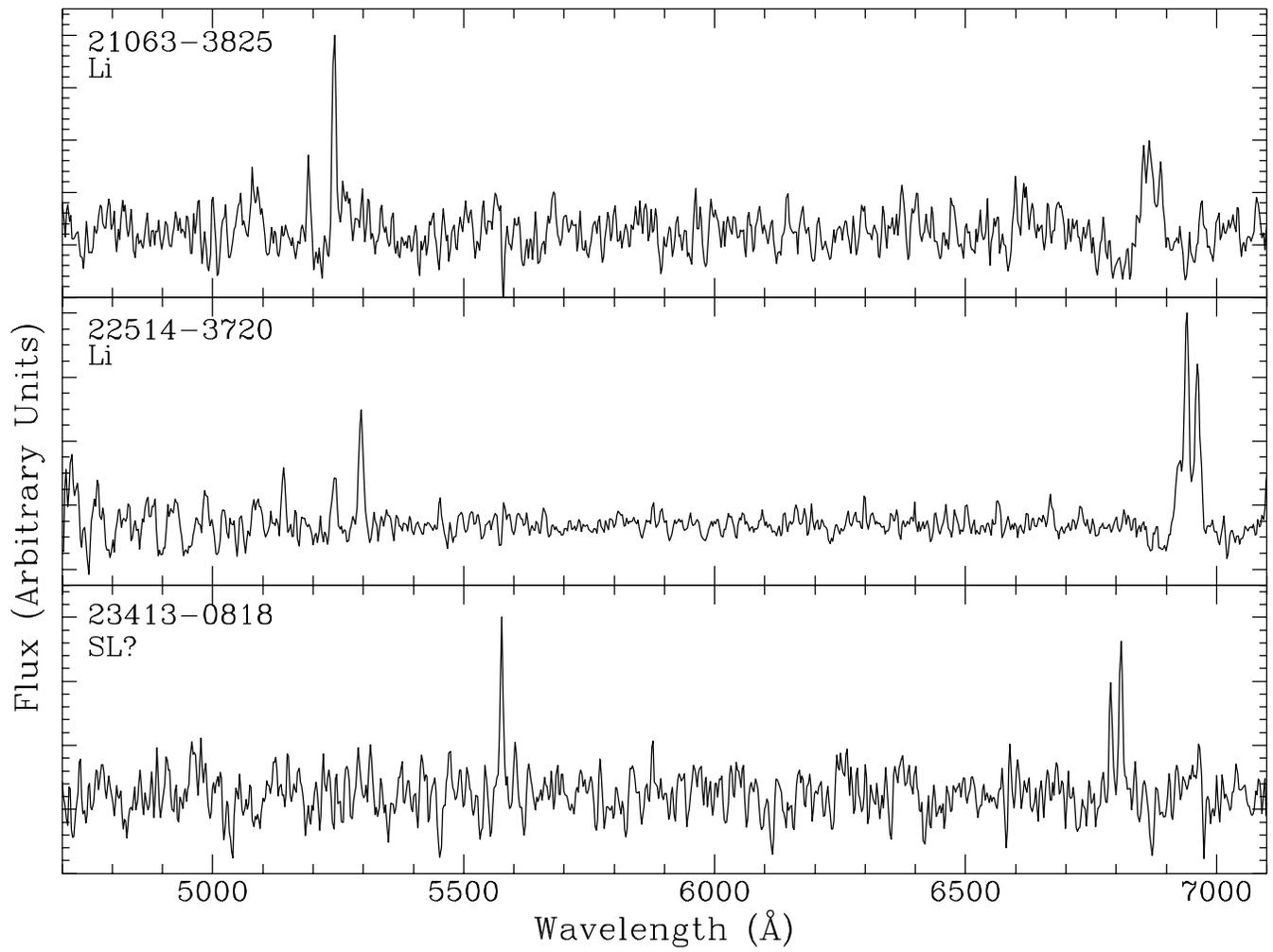

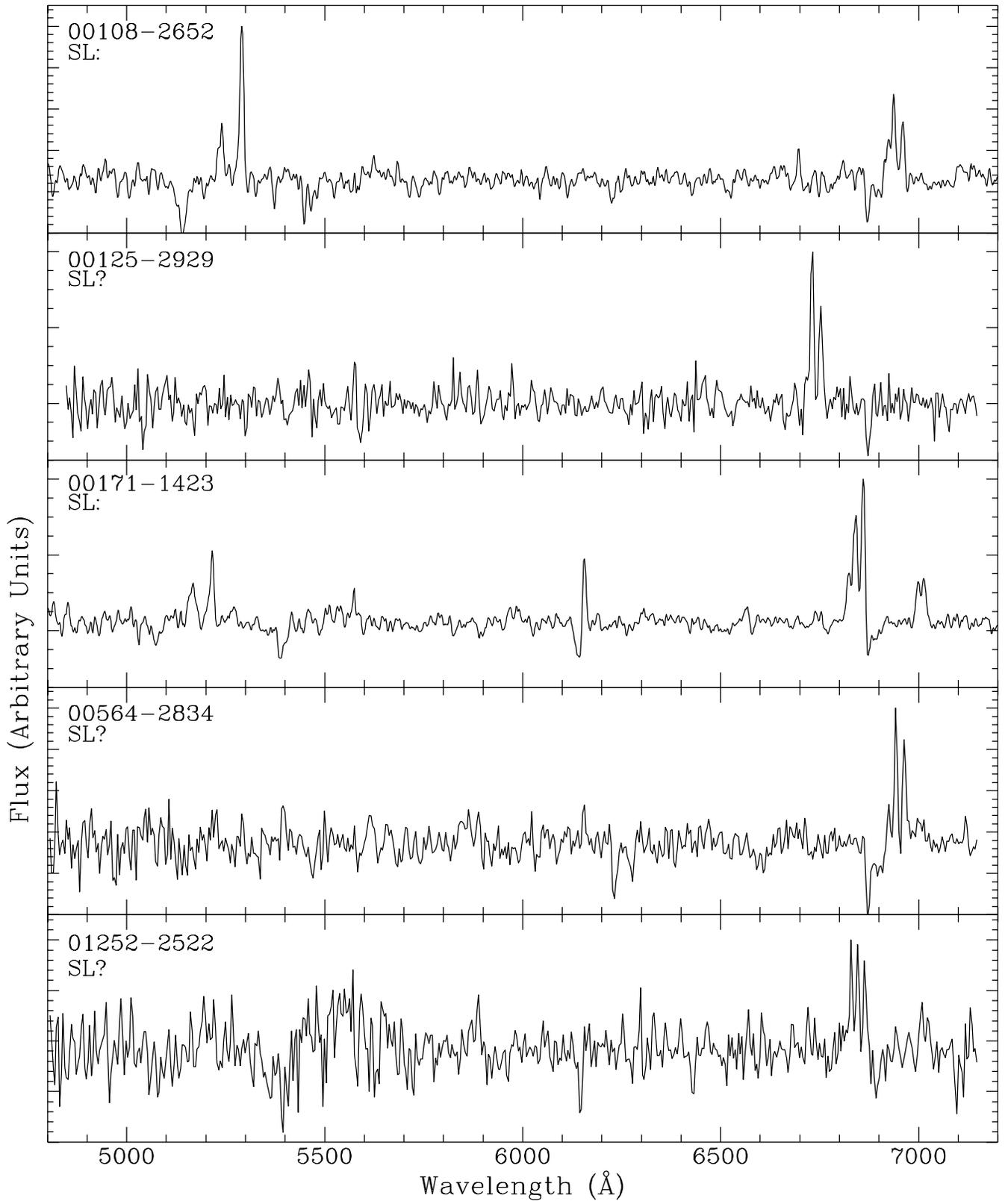

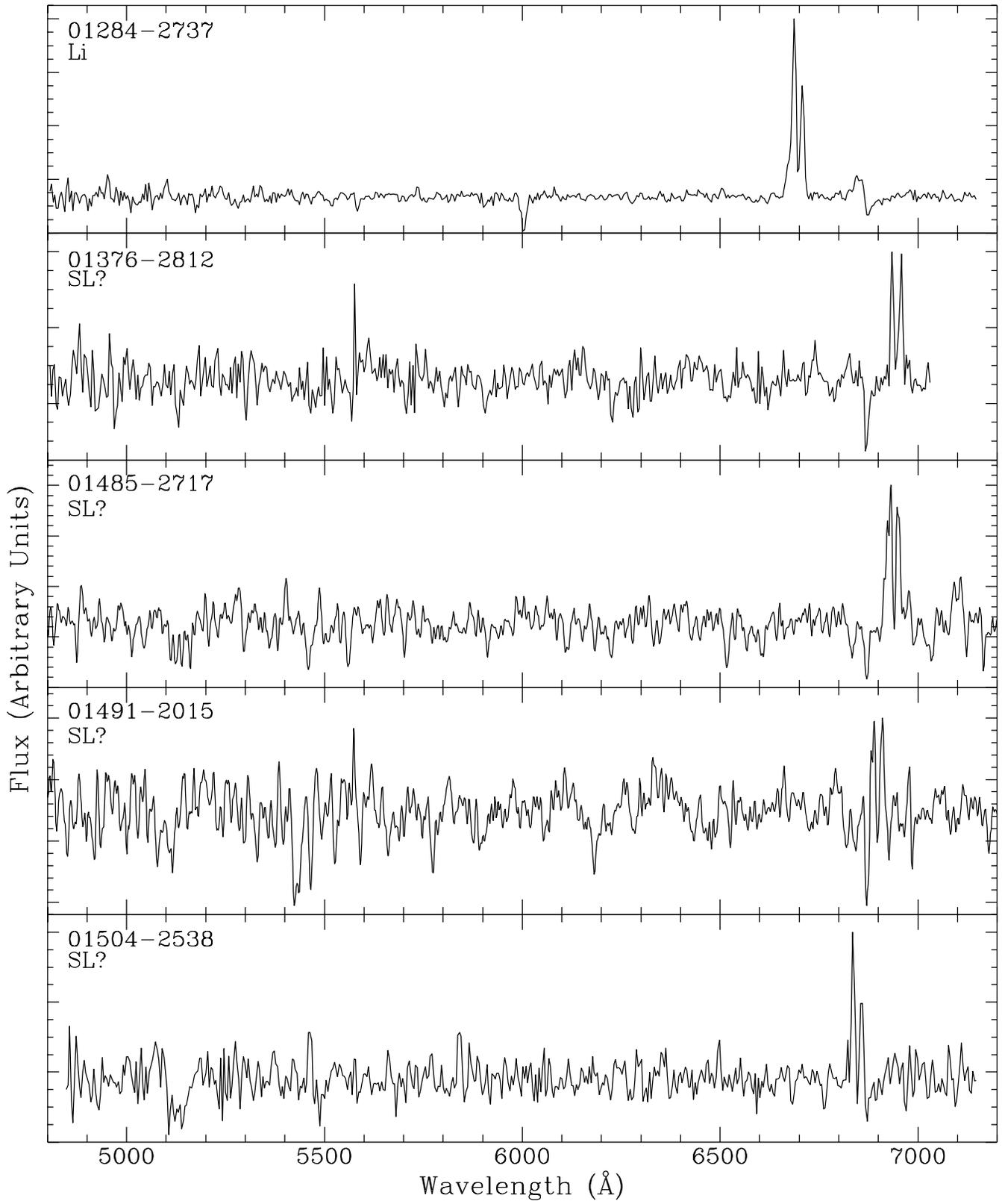

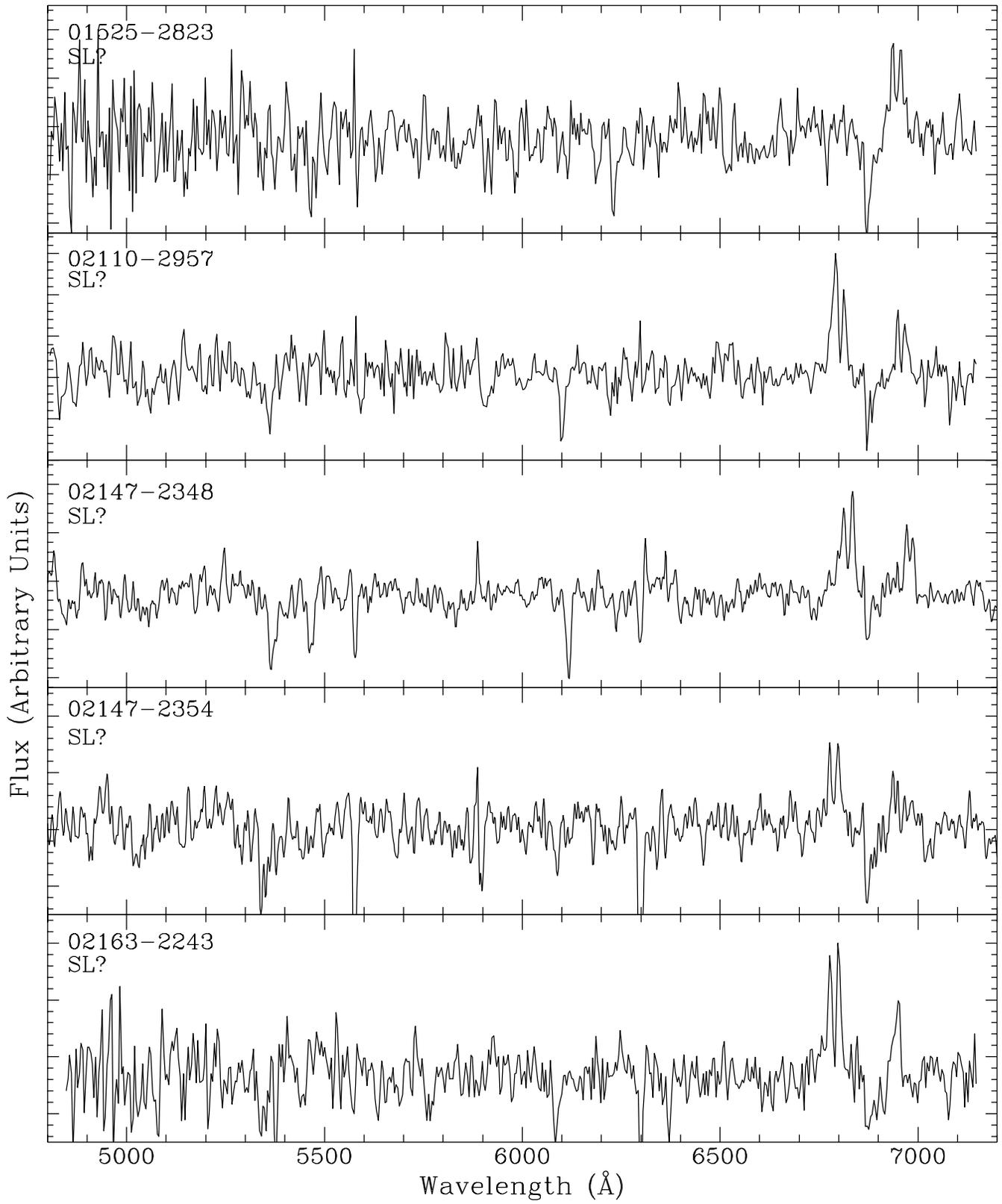

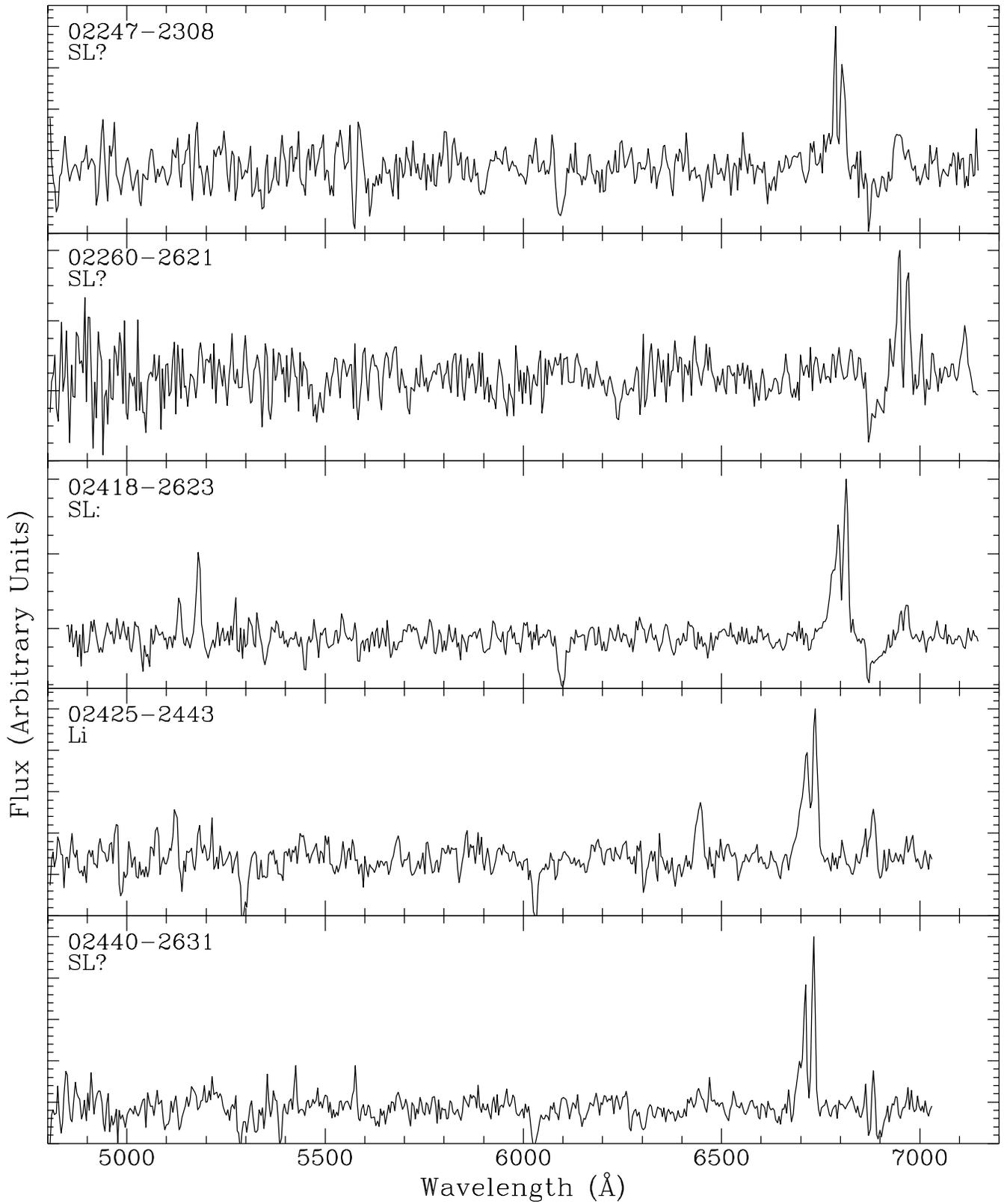

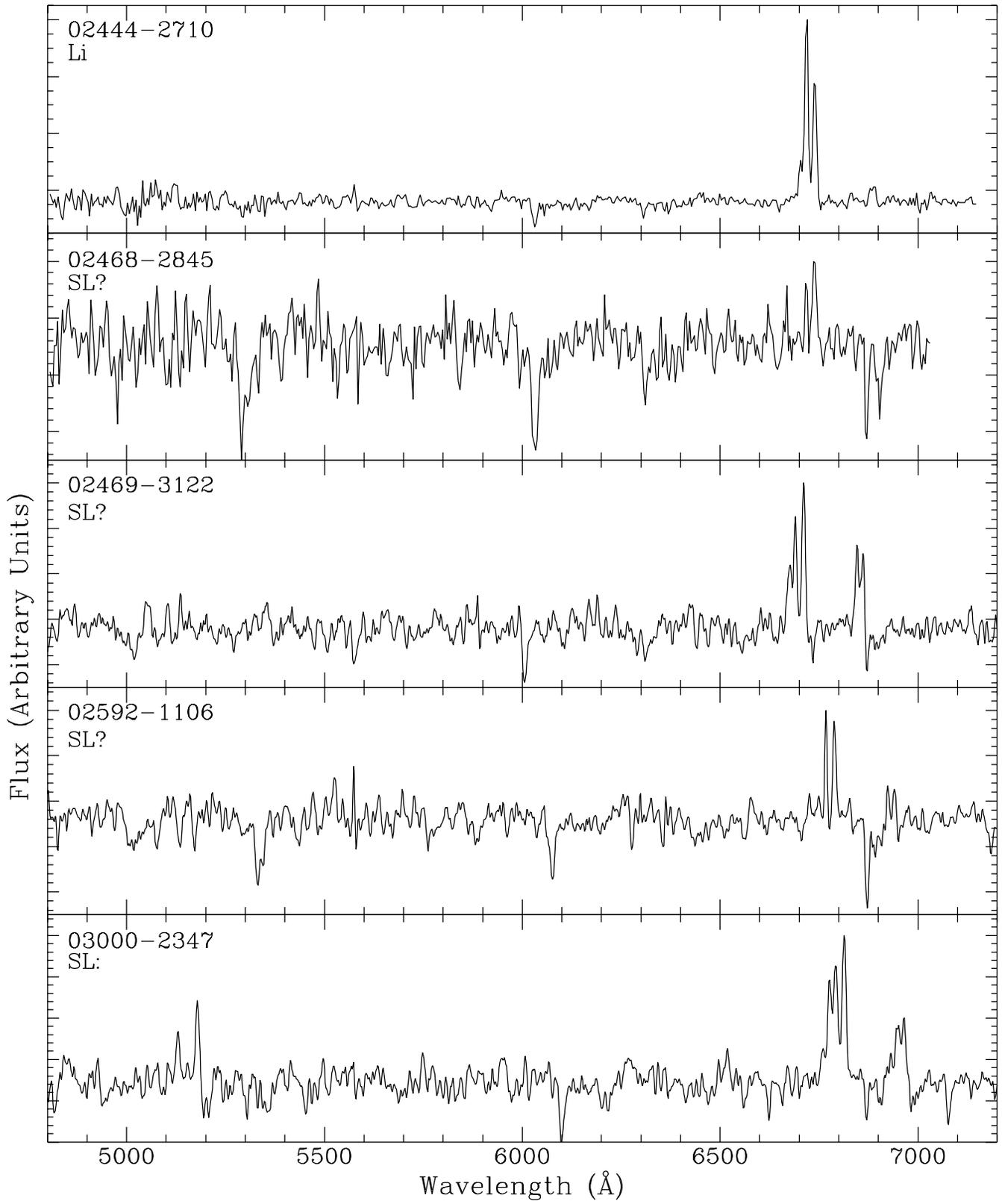

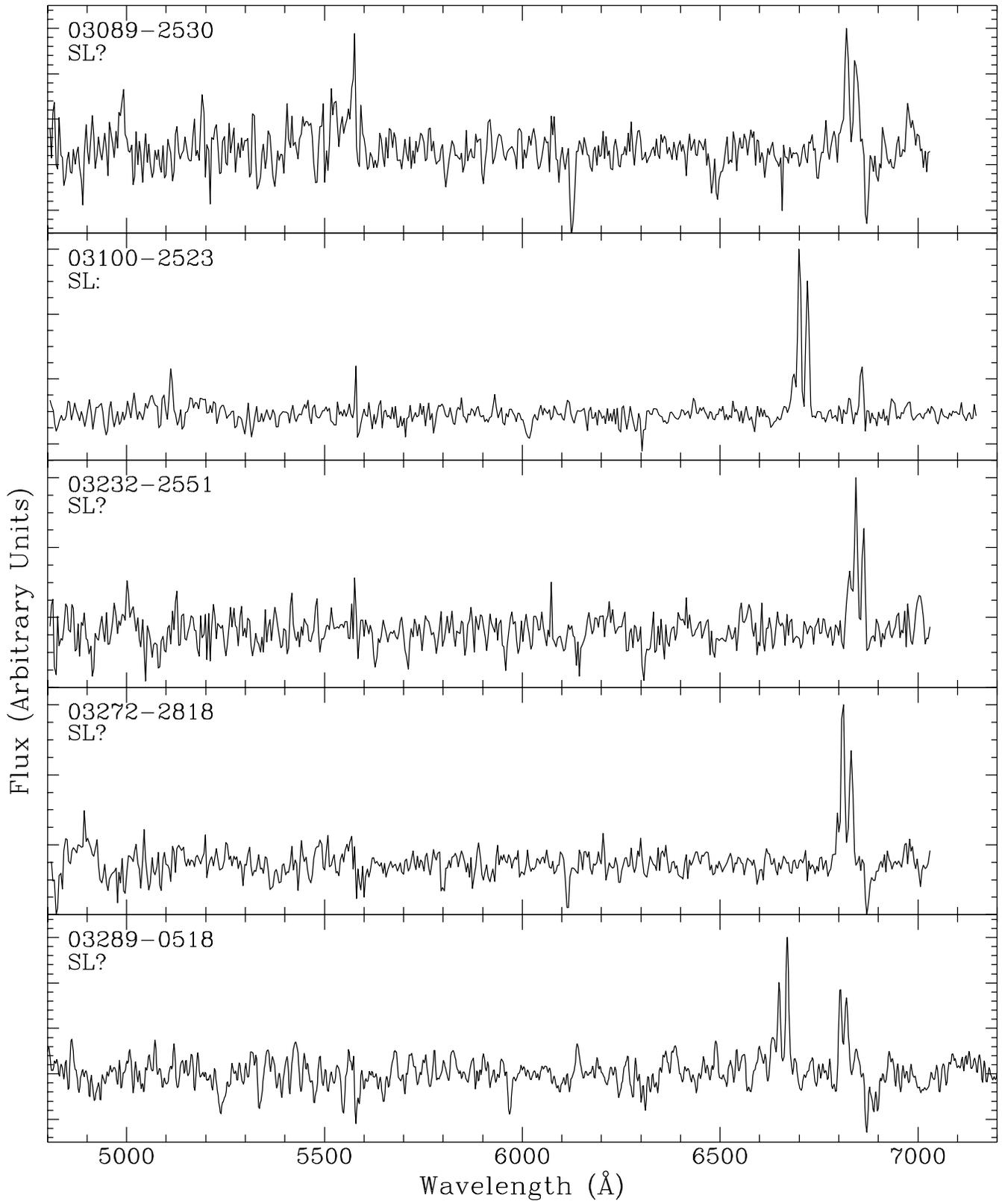

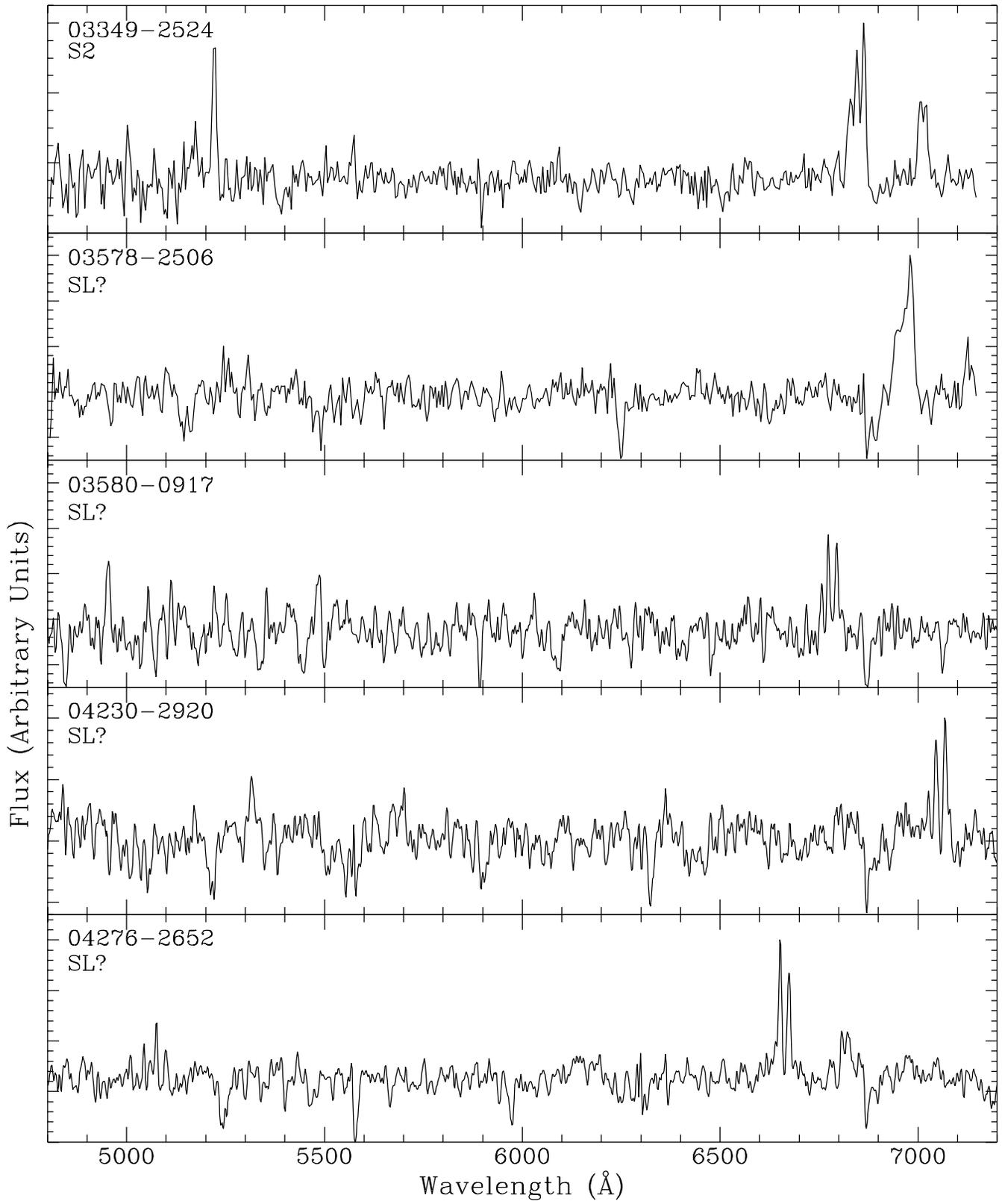

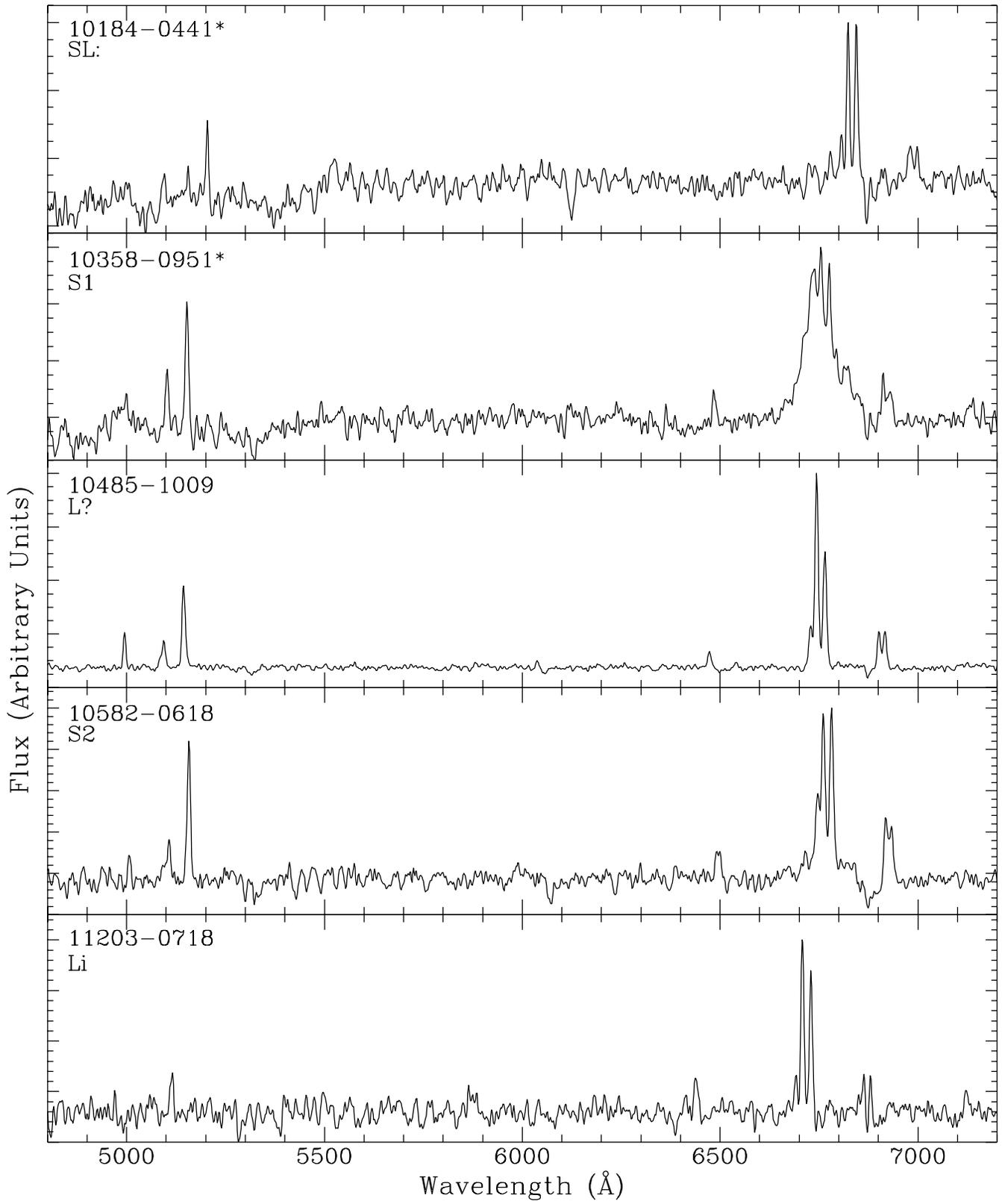

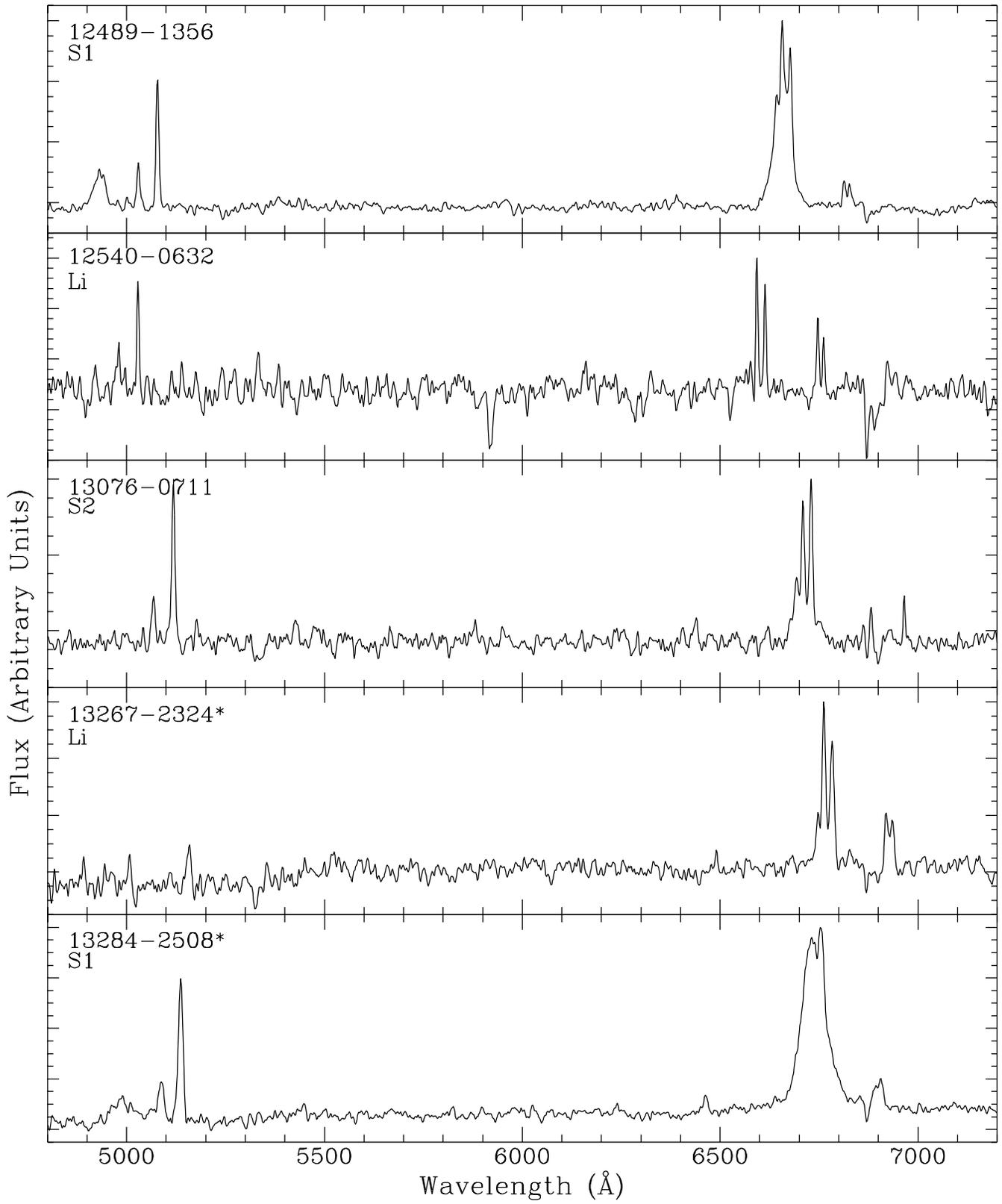

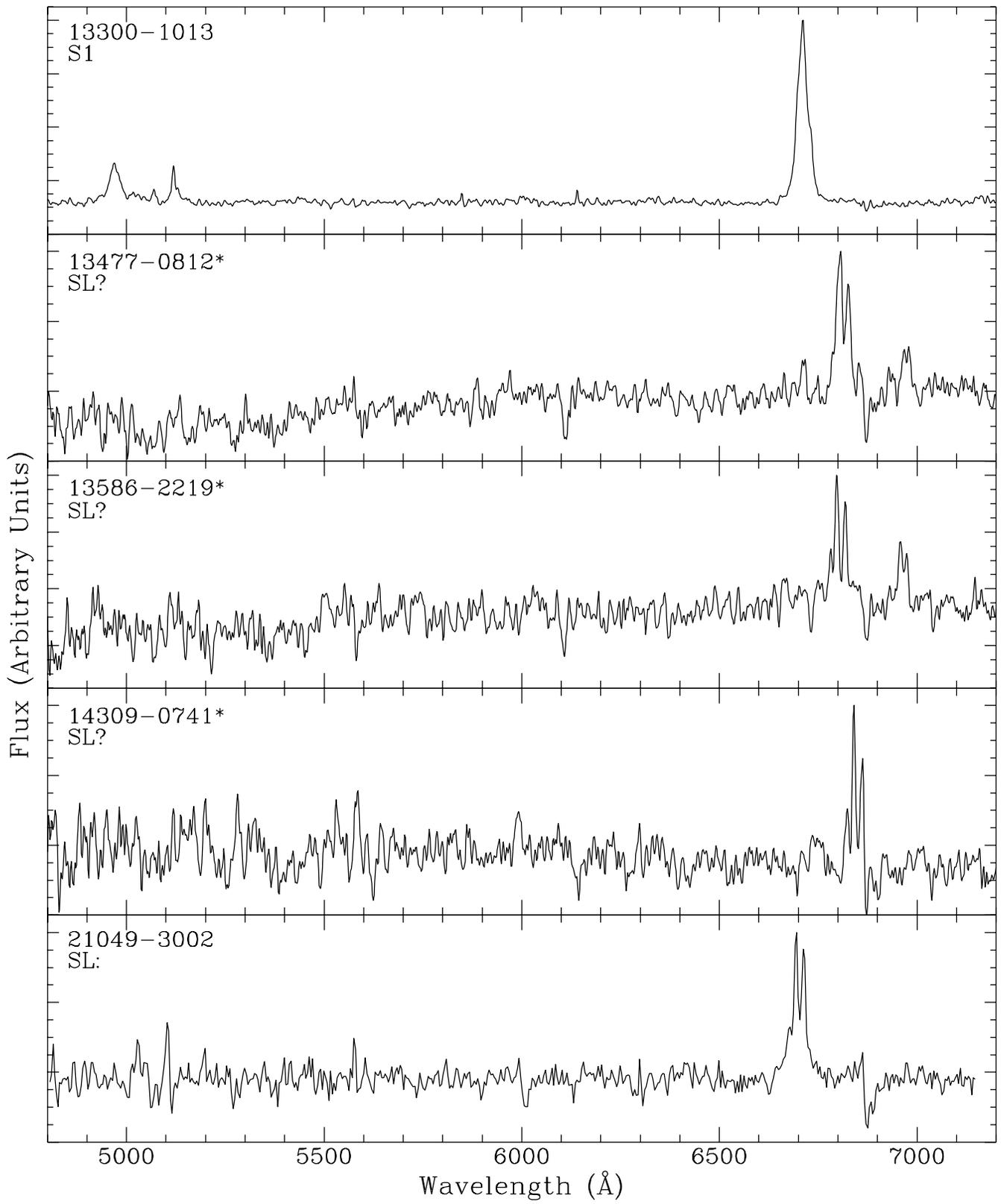

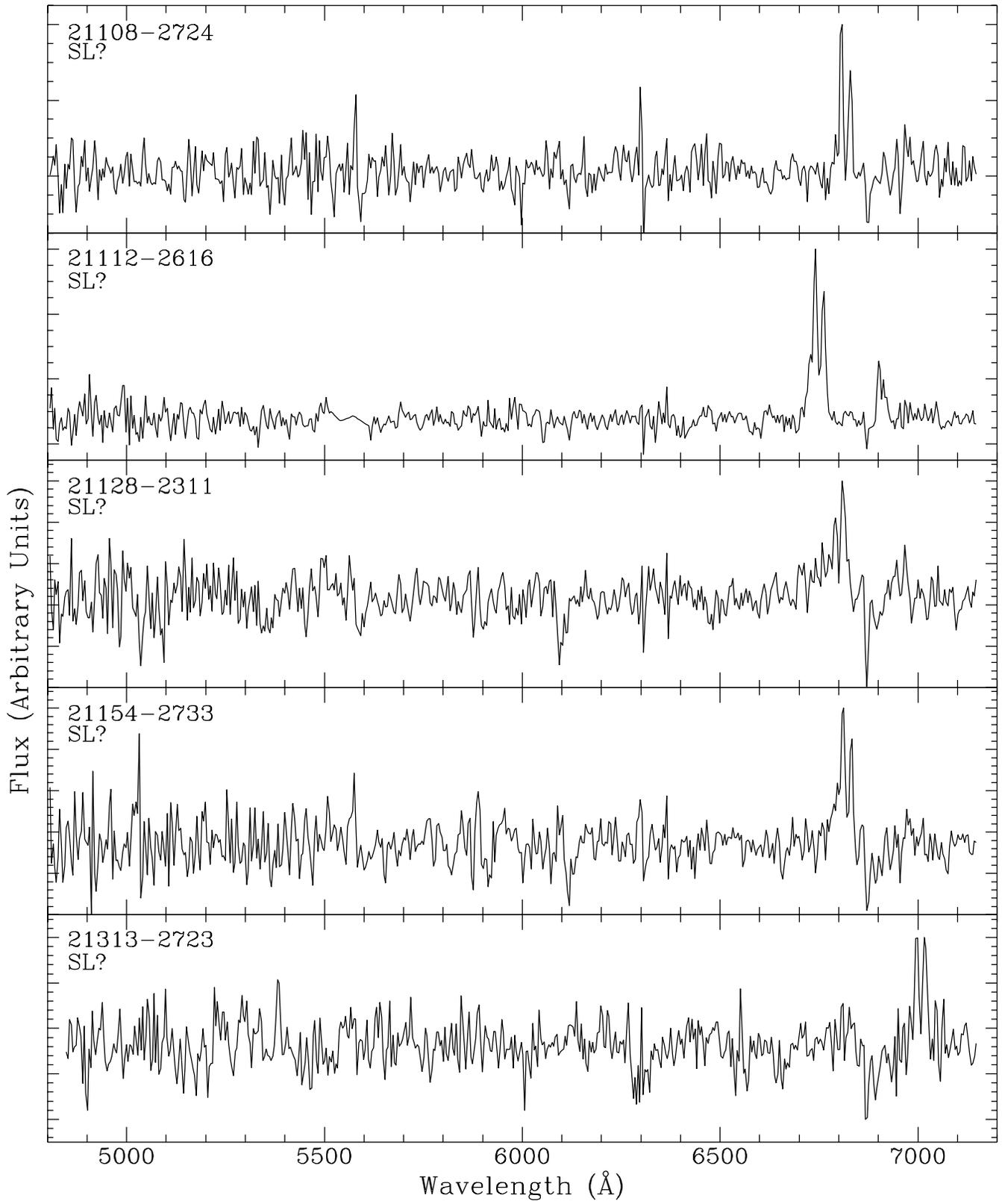

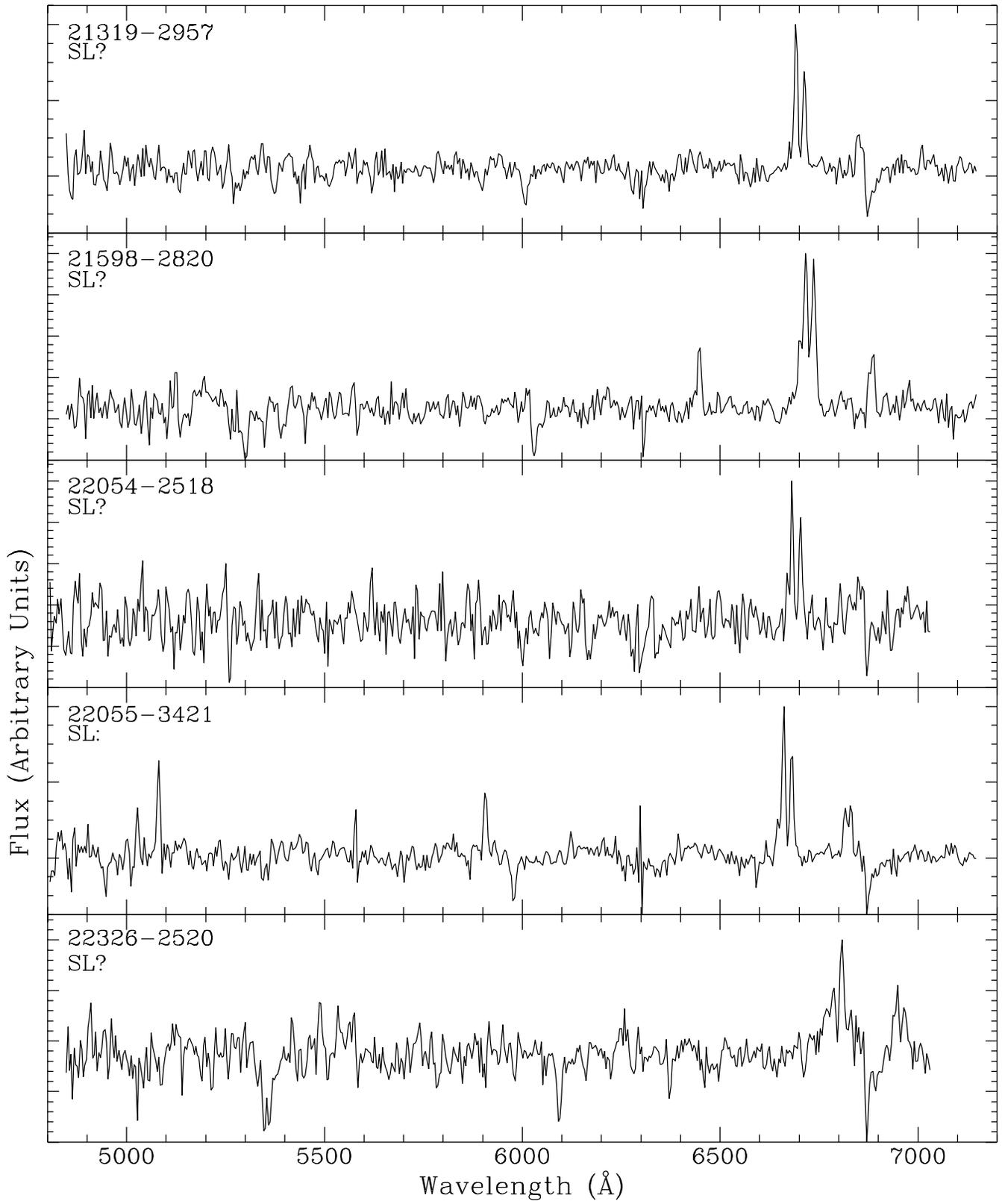

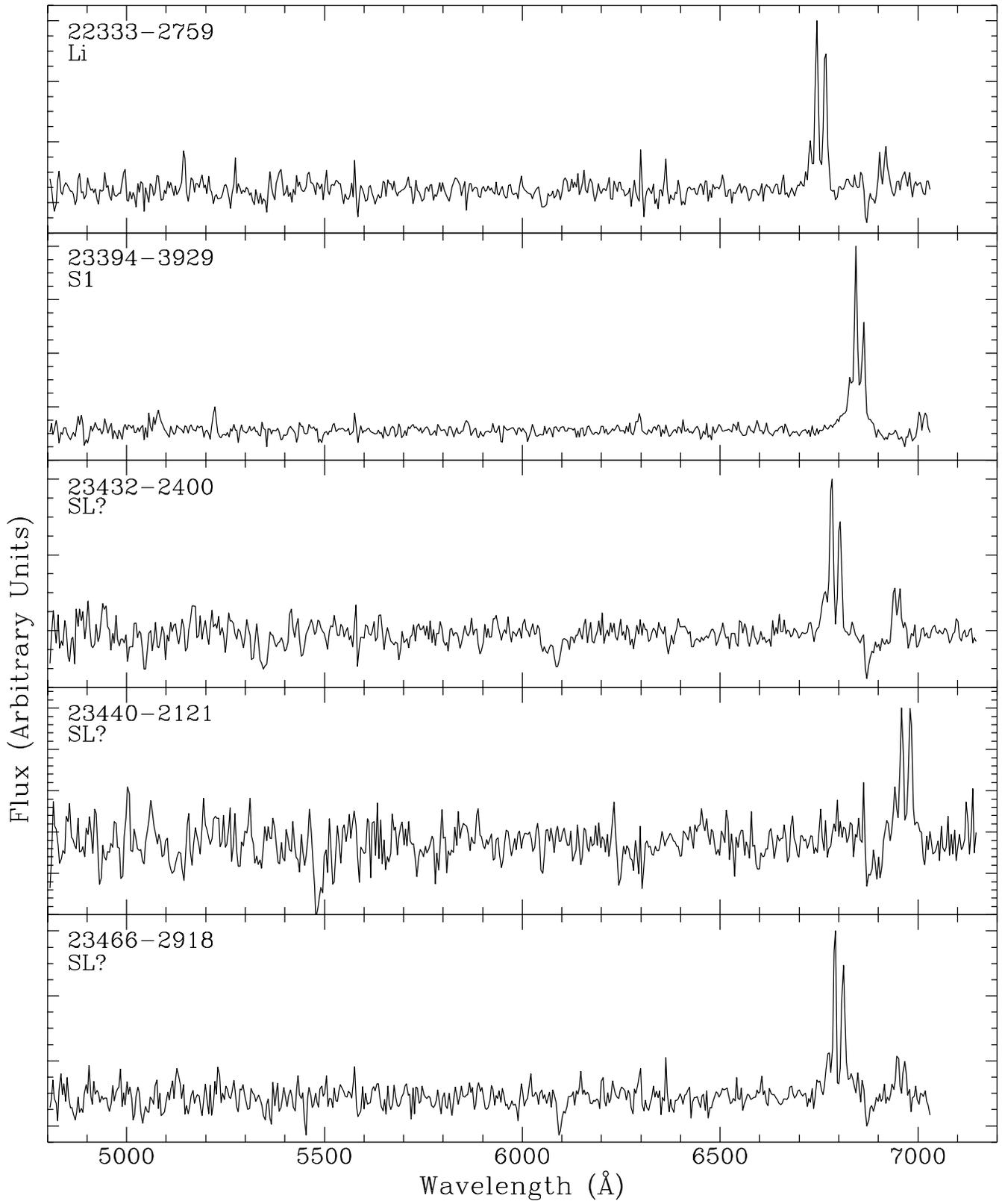

**Table 1.** List of AGN candidates observed at LNA and CASLEO.

| Catalogue Name (1) | Other Name (2) | RA (3) | Dec (1950) (4) | $m_B$ (5) | ref. (6) | Type (7) | Obs. (8) | $V_\odot$ km/s (9) | $M_B$ (10) | $log(L_{IR})$ $L_\odot$ (11) | $R_1$ (12) | $R_2$ (13) | Emiss. Type (14) | Comments (15) |
|---|---|---|---|---|---|---|---|---|---|---|---|---|---|---|
| 00283-0929 | MCG-02-02-035 | 00 28 21.4 | -09 29 02 | 15.28 | 2 | S0 * | cas | 5940 | -17.9 | | 3.44 | | SL: | 1,2,4 |
| 00539-3531 | ESO 351-G23 | 00 53 57.0 | -35 31 00 | 16.06 | 1 | Sa | cas | 14494 | -19.1 | 9.67 | 1.56 | | SL: | 1,2 |
| 00560-3655 | ESO 351-G25 | 00 56 01.0 | -36 55 48 | 15.28 | 1 | Sb | cas | 10416 | -19.2 | | 1.58 | | SL: | 3 |
| 01169-1557 | MCG-03-04-046 | 01 16 57.6 | -15 57 44 | 15.26 | 2 | SBb * | lna | 15234 | -20.0 | 10.21 | 1.42 | 12.43 | S2 | 3 |
| 01220-2422 | MCG-04-04-007 | 01 22 01.6 | -24 22 25 | 15.22 | 2 | SB(s)b | cas | 11196 | -19.4 | 10.10 | 0.84 | | SL? | 2 |
| 01369-0939 | NGC 640 | 01 36 55.7 | -09 39 15 | 15.45 | 2 | S...* | cas | 7490 | -18.3 | 9.48 | 1.27 | 28.62 | S2 | 2 |
| 02077-0917 | | 02 07 43.5 | -09 17 43 | 15.23 | 2 | Sa * | cas | 12492 | -19.6 | 9.99 | 0.93 | 3.37 | S2 | 2,4 |
| 02406-0859 | NGC 1071 | 02 40 40.8 | -08 59 08 | 15.41 | 2 | SB(rs)a | cas | 11303 | -19.2 | | 0.90 | | SL: | 2,4 |
| 02568-3508 | ESO 356-G23 | 02 56 48.0 | -35 08 54 | 15.49 | 1 | SB0-a | cas | 11010 | -19.1 | | 1.89 | | SL: | 2 |
| 03414-3154 | | 03 41 27.1 | -31 54 04 | 14.90 | 2 | S...* | cas | 9515 | -19.4 | 9.30 | 0.76 | 3.54 | S2 | 3,4 |
| 04119-3207 | ESO 420-G13 | 04 11 54.0 | -32 07 54 | 13.31 | 1 | S0(r) | cas | 3618 | -18.8 | 10.26 | 0.80 | 3.31 | S2 | 2 |
| 04303-3030 | ESO 421-G09 | 04 30 20.0 | -30 30 18 | 15.28 | 1 | S(r)a | cas | 16481 | -20.2 | 10.35 | 0.63 | 2.94 | Li | 3 |
| 04392-3739 | ESO 304-G11 | 04 39 14.1 | -37 39 52 | 15.62 | 1 | Sa | cas | 12229 | -19.2 | | 0.67 | 5.76 | S2 | 3,4 |
| 04593-1614 | MCG-03-13-051 | 04 59 21.0 | -16 14 00 | | | SB(s)b | lna | 6683 | | 9.63 | 1.42 | | SL: | 2 |
| 09400-0328 | NGC 2974 | 09 40 00.0 | -03 28 00 | | | E4 | cas | 1851 | | 8.27 | 2.86 | | SL? | 1,2,4 |
| 10192-0312 | MCG+00-27-002 | 10 19 12.0 | -03 12 00 | | | E * | cas | 12245 | | | | | S1 | 2,4 |
| 10398-1724 | MCG-03-27-026 | 10 39 51.0 | -17 24 35 | | | S0 pec | cas | 6192 | | | 1.31 | 28.26 | S2 | 2 |
| 14095-2652 | ESO 511-G10 | 14 09 31.0 | -26 52 24 | 13.53 | 1 | Sb-c | lna | 6740 | -19.9 | 9.84 | 1.20 | | SL: | 2 |
| 20200-3846 | ESO 340-G22 | 20 20 01.0 | -38 46 30 | 15.76 | 1 | SBa | lna | 16808 | -19.7 | 9.96 | 0.79 | 3.57 | S2 | 2 |
| 20211-2337 | ESO 528-G01 | 20 21 10.0 | -23 37 36 | 16.14 | 1 | Sc | cas | 16552 | -19.3 | | 0.94 | 4.65 | S2 | 2 |
| 21063-3825 | | 21 06 18.9 | -38 25 23 | 14.87 | 2 | Sa * | cas | 13934 | -20.2 | 9.94 | 1.27 | 2.01 | Li | 2 |
| 22514-3720 | ESO 406-G18 | 22 51 27.0 | -37 20 54 | 15.17 | 1 | S0 | cas | 17157 | -20.4 | 10.32 | 0.92 | 1.14 | Li | 2 |
| 23413-0818 | MCG-01-60-025 | 23 41 18.0 | -08 18 42 | 15.05 | 2 | SBb * | cas | 10337 | -19.4 | | 1.63 | | SL? | 1,2,4 |

**Table 2.** List of AGN candidates observed at ESO and CTIO.

| Catalogue Name (1) | Other Name (2) | RA (3) | Dec (1950) (4) | $m_B$ (5) | ref. (6) | Type (7) | Obs. (8) | $V_\odot$ km/s (9) | $M_B$ (10) | $log(L_{IR})$ $L_\odot$ (11) | $R_1$ (12) | $R_2$ (13) | Emiss. Type (14) | Comments (15) |
|---|---|---|---|---|---|---|---|---|---|---|---|---|---|---|
| 00108-2652 | ESO 472-IG21 | 00 10 48.0 | -26 52 18 | 15.13 | 1 | Double | eso | 17122 | -20.4 | 8.27 | 0.73 | | SL: | 1,2 |
| 00125-2929 | ESO 410-G04 | 00 12 35.0 | -29 29 18 | 14.89 | 1 | Sb | ctio | 7399 | -18.8 | 7.56 | 0.77 | | SL? | 2,4 |
| 00171-1423 | | 00 17 11.6 | -14 23 57 | 15.42 | 2 | Sb(r) * | eso | 12622 | -19.5 | 7.93 | 1.25 | | SL: | 1,2,4 |
| 00564-2834 | | 00 56 27.6 | -28 34 23 | 15.31 | 2 | Sa * | ctio | 17709 | -20.3 | 8.18 | 0.77 | | SL? | 2 |
| 01252-2522 | | 01 25 15.3 | -25 22 50 | 15.35 | 2 | Sa * | ctio | 12656 | -19.5 | | 0.81 | | SL? | 2 |
| | | | | | | | | | | | | | | |
| 01284-2737 | ESO 413-G08 | 01 28 29.0 | -27 37 18 | 15.11 | 1 | Sa | ctio | 5612 | -18.0 | 9.70 | 0.79 | 0.90 | Li | 2 |
| 01376-2812 | ESO 413-G15 | 01 37 38.0 | -28 12 18 | 15.65 | 1 | Sb | ctio | 16825 | -19.8 | 8.46 | 1.27 | | SL? | 2,4 |
| 01485-2717 | | 01 48 35.0 | -27 17 10 | 15.45 | 2 | Sab * | eso | 16835 | -20.0 | | 0.83 | | SL? | 1,2 |
| 01491-2015 | | 01 49 07.0 | -20 15 26 | 15.42 | 2 | Sa * | eso | 14865 | -19.8 | 7.60 | 1.02 | | SL? | 1,2,4 |
| 01504-2538 | | 01 50 27.4 | -25 38 26 | 15.17 | 2 | Sa * | ctio | 12486 | -17.2 | | 0.98 | | SL? | 2 |
| | | | | | | | | | | | | | | |
| 01525-2823 | ESO 414-IG09N | 01 52 32.0 | -28 23 38 | 16.49 | 1 | Double | ctio | 16945 | -19.0 | 10.21 | 0.94 | | SL? | 3,4 |
| 02110-2957 | ESO 415-G07 | 02 11 04.0 | -29 57 27 | 15.04 | 1 | S0 | ctio | 10368 | -19.4 | 7.58 | 0.83 | | SL? | 2 |
| 02147-2348 | ESO 478-G20 | 02 14 46.7 | -23 48 24 | 15.73 | 1 | Sa | eso | 11403 | -18.9 | | 1.00 | | SL? | 1,3,4 |
| 02147-2354 | | 02 14 47.4 | -23 54 45 | 15.24 | 2 | S0-a * | eso | 9818 | -19.1 | | 1.09 | | SL? | 1,2,4 |
| 02163-2243 | ESO 478-G24 | 02 16 22.0 | -22 43 30 | 15.97 | 1 | S0-a | ctio | 9835 | -18.4 | | 1.07 | | SL? | 3 |
| | | | | | | | | | | | | | | |
| 02247-2308 | | 02 24 43.3 | -23 08 09 | 15.02 | 2 | Sa * | ctio | 10178 | -19.4 | | 1.20 | | SL? | 2 |
| 02260-2621 | | 02 26 00.4 | -26 21 18 | 15.39 | 2 | Sb * | ctio | 17483 | -20.2 | | 0.89 | | SL? | 2 |
| 02418-2623 | ESO 479-G30 | 02 41 51.0 | -26 23 54 | 15.48 | 1 | S0 | ctio | 10532 | -19.0 | 7.76 | 1.46 | | SL: | 1,2 |
| 02425-2443 | ESO 479-G31 | 02 42 34.0 | -24 43 30 | 14.76 | 1 | S0 | ctio | 7053 | -18.9 | | 2.04 | 1.89 | Li | 1,3 |
| 02440-2631 | ESO 479-G35 | 02 44 01.0 | -26 31 00 | 14.79 | 1 | S0 | ctio | 6760 | -18.7 | 9.14 | 1.72 | | SL? | 2 |
| | | | | | | | | | | | | | | |
| 02444-2710 | ESO 479-G39 | 02 44 26.0 | -27 10 54 | 14.73 | 1 | Sb | ctio | 7109 | -18.9 | 7.89 | 0.74 | 1.48 | Li | 2 |
| 02468-2845 | ESO 416-G26 | 02 46 48.0 | -28 45 00 | 14.58 | 1 | S... | ctio | 6838 | -18.8 | | 2.18 | | SL? | 1 |
| 02469-3122 | ESO 416-G28 | 02 46 58.0 | -31 22 47 | 14.27 | 1 | Sc | eso | 5838 | -18.9 | 9.44 | 1.66 | | SL? | 2 |
| 02592-1106 | | 02 59 16.8 | -11 06 28 | 15.46 | 2 | Sa-b * | eso | 9337 | -18.7 | 9.48 | 0.93 | | SL? | 1,2,4 |
| 03000-2347 | | 03 00 00.2 | -23 47 05 | 14.93 | 2 | S0-a * | eso | 10515 | -19.5 | 9.92 | 1.11 | | SL: | 3 |

**Table 2.** Continued.

| Catalogue Name (1) | Other Name (2) | RA (1950) (3) | Dec (1950) (4) | $m_B$ (5) | ref. (6) | Type (7) | Obs. (8) | $V_\odot$ km/s (9) | $M_B$ (10) | $log(L_{IR})$ $L_\odot$ (11) | $R_1$ (12) | $R_2$ (13) | Emiss. Type (14) | Comments (15) |
|---|---|---|---|---|---|---|---|---|---|---|---|---|---|---|
| 03089-2530 | ESO 481-G04 | 03 08 59.0 | -25 30 36 | 15.26 | 1 | Sa | ctio | 11686 | -19.4 | | 0.81 | | SL? | 2,4 |
| 03100-2523 | | 03 10 03.6 | -25 23 26 | 15.35 | 2 | Sa * | ctio | 6482 | -18.1 | 9.38 | 0.82 | | SL: | 2,4 |
| 03232-2551 | ESO 481-G26 | 03 23 15.0 | -25 51 18 | 15.57 | 1 | Sb | ctio | 12587 | -19.3 | 7.95 | 0.83 | | SL? | 1,2 |
| 03272-2818 | ESO 418-G06 | 03 27 15.0 | -28 18 12 | 14.83 | 1 | Sb | ctio | 11415 | -19.8 | 7.90 | 0.70 | | SL? | 1,2 |
| 03289-0518 | | 03 28 54.1 | -05 18 40 | 14.65 | 2 | Sb * | ctio | 3937 | -17.7 | | 1.97 | | SL? | 1,3 |
| 03349-2524 | ESO 482-G14 | 03 34 55.0 | -25 24 48 | 15.54 | 1 | S(r)a | ctio | 13103 | -19.4 | 7.73 | 1.29 | 3.29 | S2 | 3 |
| 03578-2506 | | 03 57 48.5 | -25 06 15 | 15.47 | 2 | Sa * | ctio | 18213 | -20.2 | | | | SL? | 1,2 |
| 03580-0917 | | 03 58 01.2 | -09 17 35 | 15.45 | 2 | Sb-c * | eso | 9654 | -18.8 | | 1.12 | | SL? | 2 |
| 04230-2920 | | 04 23 01.0 | -29 20 54 | 15.39 | 2 | Sb-c * | eso | 22080 | -20.7 | | 1.16 | | SL? | 1,2 |
| 04276-2652 | ESO 484-G26 | 04 27 39.0 | -26 52 48 | 14.96 | 1 | Sa | eso | 4099 | -17.5 | 10.97 | 0.93 | | SL? | 2,4 |
| 10184-0441 | | 10 18 26.0 | -04 41 54 | | | Sb * | eso | 11890 | | | 1.01 | | SL: | 1,2 |
| 10358-0951 | | 10 35 51.0 | -09 51 23 | | | SBb * | eso | 8710 | | | | 2.85 | S1 | 2 |
| 10485-1009 | | 10 48 30.0 | -10 09 27 | | | SBa-b * | eso | 8245 | | 7.69 | 0.66 | 1.76 | L? | 2 |
| 10582-0618 | | 10 58 14.0 | -06 18 35 | | | Sb * | eso | 8998 | | 7.59 | 1.08 | 3.83 | S2 | 3 |
| 11203-0718 | | 11 20 19.0 | -07 18 51 | | | Sa(r) * | eso | 6628 | | | 0.77 | 2.07 | Li | 3 |
| 12489-1356 | | 12 48 54.0 | -13 56 57 | | | Sa * | eso | 4298 | | 9.28 | 1.03 | 0.85 | S1 | 2 |
| 12540-0632 | | 12 54 00.0 | -06 32 51 | | | Sa * | eso | 1373 | | | 0.84 | 2.02 | Li | 3 |
| 13076-0711 | | 13 07 41.0 | -07 11 18 | | | S...* | eso | 6713 | | 9.56 | 1.17 | 21.37 | S2 | 3 |
| 13267-2324 | ESO 509-G29 | 13 26 45.0 | -23 24 30 | 15.39 | 1 | N | eso | 9144 | -18.8 | 8.06 | 0.89 | 1.70 | Li | 2 |
| 13284-2508 | ESO 509-G38 | 13 28 28.0 | -25 08 42 | 14.75 | 1 | S... | eso | 7787 | -19.1 | 9.66 | | | S1 | 2 |
| 13300-1013 | | 13 30 00.0 | -10 13 29 | | | SBb * | eso | 6639 | | | 0.53 | 0.32 | S1 | 2 |
| 13477-0812 | | 13 47 42.0 | -08 12 29 | | | Sb-c * | eso | 11089 | | | 0.91 | | SL? | 1,2 |
| 13586-2219 | ESO 578-G15 | 13 58 38.0 | -22 19 48 | 15.25 | 1 | Sb | eso | 10828 | -19.3 | | 0.83 | | SL? | 1,2 |
| 14309-0741 | | 14 30 56.0 | -07 41 53 | | | E * | eso | 12682 | | | 0.75 | | SL? | 1,2,4 |
| 21049-3002 | | 21 04 59.7 | -30 02 17 | 15.10 | 2 | S0 * | ctio | 5751 | -18.1 | 7.23 | 0.99 | | SL: | 2,4 |

**Table 2.** Continued.

| Catalogue Name (1) | Other Name (2) | RA (3) | Dec (1950) (4) | $m_B$ (5) | ref. (6) | Type (7) | Obs. (8) | $V_\odot$ km/s (9) | $M_B$ (10) | $log(L_{IR})$ $L_\odot$ (11) | $R_1$ (12) | $R_2$ (13) | Emiss. Type (14) | Comments (15) |
|---|---|---|---|---|---|---|---|---|---|---|---|---|---|---|
| 21108-2724 |  | 21 10 50.3 | -27 24 30 | 15.39 | 2 | Sb-c * | ctio | 10878 | -19.2 |  | 0.70 |  | SL? | 2,4 |
| 21112-2616 | ESO 530-G25 | 21 11 14.0 | -26 16 06 | 15.35 | 1 | Sa | ctio | 8051 | -18.5 | 9.72 | 0.83 |  | SL? | 2 |
| 21128-2311 |  | 21 12 51.6 | -23 11 55 | 15.41 | 2 | S...* | ctio | 10439 | -19.0 |  | 1.92 |  | SL? | 1 |
| 21154-2733 | ESO 464-G31N | 21 15 29.0 | -27 33 41 |  |  | Sa? pec | ctio | 11084 |  | 10.50 | 0.70 |  | SL? | 4 |
| 21313-2723 |  | 21 31 23.4 | -27 23 29 | 14.93 | 2 | S...* | ctio | 19695 | -20.9 |  | 1.02 |  | SL? | 2 |
| 21319-2957 |  | 21 31 59.6 | -29 57 13 | 14.95 | 2 | S0 * | ctio | 6048 | -18.3 | 9.22 | 0.73 |  | SL? |  |
| 21598-2820 |  | 21 59 50.8 | -28 20 07 | 15.18 | 2 | S0 * | ctio | 7211 | -18.5 |  | 0.99 |  | SL? | 3 |
| 22054-2518 | ESO 532-G21 | 22 05 25.0 | -25 18 24 | 14.65 | 1 | Sb | ctio | 5558 | -18.4 | 8.96 | 0.89 |  | SL? | 2 |
| 22055-3421 | ESO 404-G32 | 22 05 33.0 | -34 21 06 | 15.17 | 1 | S... | ctio | 4431 | -17.4 | 9.26 | 0.79 |  | SL: | 4 |
| 22326-2520 | ESO 533-G51 | 22 32 41.0 | -25 20 06 | 14.54 | 1 | Sa | ctio | 10039 | -19.8 |  | 1.18 |  | SL? | 1 |
| 22333-2759 |  | 22 33 20.6 | -27 59 03 | 14.98 | 2 | Sa-b * | ctio | 8113 | -18.9 | 9.45 | 0.93 | 1.23 | L1 | 2,4 |
| 23394-3929 |  | 23 39 26.9 | -39 29 37 | 15.42 | 2 | SBb * | ctio | 12795 | -19.5 |  | 0.97 |  | S1 | 2 |
| 23432-2400 |  | 23 43 16.7 | -24 00 33 | 15.48 | 2 | Sb-c * | ctio | 9984 | -18.9 | 8.38 | 0.82 |  | SL? | 2 |
| 23440-2121 |  | 23 44 03.1 | -21 21 17 | 15.44 | 2 | Sb(r) * | ctio | 18101 | -20.2 |  | 1.04 |  | SL? | 2 |
| 23466-2918 | ESO 471-G23 | 23 46 40.0 | -29 18 30 | 14.97 | 1 | Sa | ctio | 10308 | -19.5 | 9.65 | 1.02 |  | SL? | 2 |